\documentclass[journal]{IEEEtran}

\ifCLASSINFOpdf
\else
   \usepackage[dvips]{graphicx}
\fi
\usepackage{url}

\usepackage{color}
\usepackage{amsmath}
\usepackage{xcolor}
\usepackage{graphicx}
\usepackage{booktabs}
\usepackage{url}
\usepackage{enumitem}
\usepackage{cite}

\begin{document}

\title{Transferable Adversarial Attacks against ASR}

\author{Xiaoxue Gao, \IEEEmembership{Member, IEEE}, Zexin Li, Yiming Chen, Cong Liu and Haizhou Li, \IEEEmembership{Fellow, IEEE}
\thanks{ Xiaoxue Gao is with the Institute for Infocomm Research, Agency for Science, Technology, and Research (A*STAR), Singapore 138632 (e-mail: Gao\textunderscore Xiaoxue@i2r.a-star.edu.sg). Yiming Chen is with National University of Singapore, Singapore 117583 (e-mail: yiming.chen@u.nus.edu). Zexin Li and Cong Liu are with University of California Riverside, CA, US 92521. (e-mails: zli536@ucr.edu and congl@ucr.edu). Zexin Li is the corresponding author. Haizhou Li is with Shenzhen Research Institute of Big Data, The Chinese University of Hong Kong, Shenzhen, Guangdong, 518172, P.R (e-mail: haizhouli@cuhk.edu.cn). The research is supported by National Natural Science Foundation of China (Grant No. 62271432), Shenzhen Science and Technology Program ZDSYS20230626091302006, and Shenzhen Science and Technology Research Fund (Fundamental Research Key Project Grant No. JCYJ20220818103001002).}
}

\maketitle
\vspace{-0.9cm}
\begin{abstract}
Given the extensive research and real-world applications of automatic speech recognition (ASR), ensuring the robustness of ASR models against minor input perturbations becomes a crucial consideration for maintaining their effectiveness in real-time scenarios. Previous explorations into ASR model robustness have predominantly revolved around evaluating accuracy on white-box settings with full access to ASR models. Nevertheless, full ASR model details are often not available in real-world applications. Therefore, evaluating the robustness of black-box ASR models is essential for a comprehensive understanding of ASR model resilience. In this regard, we thoroughly study the vulnerability of practical black-box attacks in cutting-edge ASR models and propose to employ two advanced time-domain-based transferable attacks alongside our differentiable feature extractor. We also propose a speech-aware gradient optimization approach (SAGO) for ASR, which forces mistranscription with minimal impact on human imperceptibility through voice activity detection rule and a speech-aware gradient-oriented optimizer. Our comprehensive experimental results reveal performance enhancements compared to baseline approaches across five models on two databases. 
\end{abstract}


\IEEEpeerreviewmaketitle

\vspace{-0.5cm}
\section{Introduction}

\IEEEPARstart{A}{utomatic} speech recognition (ASR) aims to recognize text from speech signals. ASR represents a dynamic and rapidly evolving research domain, striving to bridge the cognitive gap between human speech comprehension and computational interpretation. This progress is driven by a multitude of real-world applications such as virtual assistants (e.g., Siri and Amazon Alexa), captioning, subtitling, healthcare \cite{daniels1985just}, and autonomous vehicle systems \cite{fagnant2015preparing}.
Consequently, attaining human-like recognition performance against minor input perturbations emerges as a critical consideration for real-time applications within the ASR domain \cite{abdullah2021hear,hussain2021waveguard,yang2019characterizing,rajaratnam2018noise}.

Recognizing the significance of probing ASR system robustness, recent researches have delved into white-box scenarios to evaluate the accuracy robustness of ASR models via targeted and untargeted adversarial attacks~\cite{carlini2017towards,olivier2022there,wang2020adversarial,wu2023kenku}. In this paper, we focus on untargeted attacks. Untargeted attacks against white-box ASR models endeavor to create adversarial samples by introducing subtle perturbations to the ASR inputs, resulting in diminished recognition performance while maintaining imperceptibility to humans \cite{carlini2018audio,yakura2019robust,olivier2022recent,schonherr2018adversarial,chen2020metamorph,qin2019imperceptible,olivier2022there}.
Although the mentioned endeavors in white-box ASR have made significant advancements, encompassing DNN \cite{schonherr2018adversarial}, RNN \cite{carlini2018audio,yakura2019robust,chen2020metamorph,qin2019imperceptible}, Transformers \cite{olivier2022recent}, and the latest state-of-the-art (SOTA) Whisper ASR models \cite{olivier2022there}, they are constrained by an impractical assumption that requires the adversary to have access to the ASR model's internal information. This limitation restricts their applicability in real-world scenarios \cite{wang2022query,abdullah2021sok}.

Adversarial attacks pose a critical challenge in security-sensitive and safety-critical domains~\cite{madry2017towards,kurakin2016adversarial}.
Various adversarial attack approaches have been extensively studied in computer vision~\cite{zhang2022investigating,li2023sibling} and natural language processing~\cite{li2023white,ebrahimi2018hotflip,li2019textbugger}. 
With advancements in white-box ASR attacks, the potential and exploration of black-box ASR attacks, especially transferable attacks, have also been gaining attention.
Notable exceptions in attacking Kaldi-based models \cite{chen2020devil} and RNN-based ASR models include universal adversarial perturbations \cite{neekhara2019universal} and study on factors affecting target transferability~\cite{abdullah2021demystifying}. 
The above transferable attacks involve adversaries using adversarial examples crafted for one trained model to target other black-box ASR models. By eliminating the need to access the full black-box model architecture and weights, transfer attacks offer greater flexibility and possibilities for real-world applications.
However, transfer attacks against advanced black-box ASR models (e.g., Whisper~\cite{radford2023robust}, Speech2text \cite{wang2020fairseq}) still remain unexplored.
The accuracy robustness of black-box ASR still lags considerably behind human speech recognition performance, highlighting the need for a thorough exploration of recent models and advanced transferable attack methods.

Recent advancements on attack transferability through innovative optimization~\cite{dong2018boosting,kurakin2016adversarial,wang2021enhancing,zhou2018transferable,zhang2022improving,huang2022transferable} offer valuable insights for our work.
To this regard, this paper aims to conduct a comprehensive exploration of accuracy robustness on cutting-edge ASR models, i.e., Whisper \cite{olivier2022there} and Transformer \cite{wang2020fairseq}.
Inspired by~\cite{dong2018boosting,wang2021enhancing}, we propose to employ momentum iterative fast gradient sign
method (MI-FGSM) and variance tuning momentum iterative fast
gradient sign method (VMI-FGSM) directly on time-domain audio signals through differentiable feature extraction design. 

On the other hand, integrating VAD tasks into ASR has proven effective in enhancing recognition accuracy~\cite{radford2023robust,li2022incorporating}. 
VAD is designed to distinguish speech from non-speech segments in audio input \cite{novitasari2022improving}.
Its close relationship with ASR lies in the fact that it identifies speech portions that are crucial for ASR~\cite{wang2022vadoi,radford2023robust,li2022incorporating}. Specifically, the correlation stems from the fact that the commencement of speech segments within an utterance usually corresponds with the detection of speech initiation by the VAD system, rendering non-speech segments identified by VAD potentially redundant for ASR \cite{radford2023robust,li2022incorporating}.
Furthermore, VAD and ASR often share optimization objectives to improve ASR system performance~\cite{radford2023robust,li2022incorporating,wang2022vadoi}. 

Motivated by this correlation, we propose a novel approach, \textbf{S}peech-\textbf{A}ware \textbf{G}radient \textbf{O}ptimization (SAGO) for black-box ASR attack.
SAGO investigates speech-aware adversarial attacks tailored to ASR, specifically targeting speech segments using a VAD mask for gradient optimization, highlighting speech parts that the human brain tends to focus on.

The contributions of this paper include: (a) \textbf{Comprehensive Exploration}: We conduct a thorough investigation into novel gradient-based strong transferable attacks on SOTA ASR models, benefiting from a differentiable feature extractor design that enables effective time-domain attacks; (b) \textbf{Innovative Methodology}: We propose an importance-attribute attack via speech-aware gradient optimization, allowing us to emulate the human brain for targeting speech segments attentively for attacks through a VAD rule; and (c) \textbf{Extensive Experimentation}: We conduct a systematic evaluation of various transferable attacks across five models on two different datasets.

\begin{table*}[t]
\vspace{-1.2cm}
\label{table1}
\caption{Comparison of the recognition performance (WER\%) of various attack approaches under ljspeech dataset. Source refers to the white-box source ASR model, while target pertains to the black-box victim ASR model. Clean and noise denote the recognition performance when decoding original clean speech and noisy speech samples, respectively. * denotes white-box attacks.}
\resizebox{\textwidth}{!}{
\begin{tabular}{ll|rrrrrrrrrr}
\toprule \midrule
\multicolumn{1}{c}{} & \textbf{Target Models} & \multicolumn{2}{c}{\textbf{Whisper-tiny}} & \multicolumn{2}{c}{\textbf{Whisper-base}} & \multicolumn{2}{c}{\textbf{S2T-small}} & \multicolumn{2}{c}{\textbf{S2T-medium}}& \multicolumn{2}{c}{\textbf{S2T-large}} \\\midrule
\multicolumn{1}{c}{\textbf{Source Models}} & \textbf{SNR}& \multicolumn{1}{c}{\textbf{30}} & \multicolumn{1}{c}{\textbf{35}} & \multicolumn{1}{c}{\textbf{30}} & \multicolumn{1}{c}{\textbf{35}} & \multicolumn{1}{c}{\textbf{30}} & \multicolumn{1}{c}{\textbf{35}} & \multicolumn{1}{c}{\textbf{30}} & \multicolumn{1}{c}{\textbf{35}} & \multicolumn{1}{c}{\textbf{30}} & \multicolumn{1}{c}{\textbf{35}} \\\midrule
& Clean & \multicolumn{2}{c}{5.32}& \multicolumn{2}{c}{3.77}& \multicolumn{2}{c}{4.77}& \multicolumn{2}{c}{4.35}& \multicolumn{2}{c}{4.54}\\
\textbf{} & White Noise \cite{olivier2022there}& 6.14&5.62  &4.49  &3.66  &3.72 & 4.90 &3.31  & 4.38 &3.07  &4.63\\
\textbf{} & Babble Noise \cite{shi2022robust}& 5.35 & 5.25 & 3.81 & 3.81 & 4.87 & 4.90 & 4.53 & 4.50 & 4.63 & 4.63 \\
 & Music Noise \cite{shi2022robust}& 5.71 & 5.12 & 3.86 & 3.74& 4.98 & 4.86& 4.40 & 4.44 & 4.63 & 4.50 \\
 & Natural Noise \cite{shi2022robust}& 5.11 & 5.04 & 3.96 & 3.77& 4.95 & 4.81& 4.62 & 4.62 &4.81 & 4.62 \\
\midrule
\textbf{Whisper-tiny} & PGD \cite{olivier2022there} & 100.00$^*$ & 93.24$^*$& 41.62& 26.75& 21.01& 13.68& 16.73& 11.75& 13.49& 10.17\\
& SAGO & 97.22$^*$&89.11$^*$& 43.48& 28.84& 22.75& 14.83& 18.95& 12.80& 15.45& 10.92\\
& VMI-FGSM & 100.00$^*$ & 85.56$^*$ & \textbf{49.03} & 31.68 & 23.98 & 16.37 & 19.97 & 14.19 & 16.96 & 11.93 \\
 & MI-FGSM & 100.00$^*$ & 89.24$^*$ & 46.97 & \textbf{31.72} & \textbf{24.48} & \textbf{17.32} & \textbf{20.13} & \textbf{14.20} & \textbf{18.12} & \textbf{12.41} \\\midrule
\textbf{Whisper-base} & PGD \cite{olivier2022there} & 49.79 & 26.75 & 100.00$^*$ & 75.88$^*$& 20.53& 14.94& 17.87& 13.25& 15.23& 11.85\\
& SAGO & 52.82& 35.69& 84.00$^*$& 65.45$^*$& 21.80& 15.93& 20.09& 14.14& 17.72& 12.69\\
& VMI-FGSM& 49.89& \textbf{38.97} & 87.31$^*$& 69.56$^*$& 23.91& 16.91& 21.98& 15.41& \textbf{23.91} & \textbf{18.80} \\
& MI-FGSM & \textbf{52.90} & 37.80& 87.00$^*$& 74.35$^*$& \textbf{24.81} & \textbf{17.33} & \textbf{22.44} & \textbf{15.94} & 19.46& 13.84\\\midrule
\textbf{S2T-small} & PGD \cite{olivier2022there} & 10.50& 9.44 & 7.40 & 6.45 & 100.00$^*$ & 100.00$^*$& 30.15& 22.38& 19.17& 15.59\\
& SAGO & 13.91& 10.77& 9.75 & 7.41 & 82.86$^*$& 63.76$^*$& \textbf{35.22} & 24.08& \textbf{23.89} & 16.30\\
& VMI-FGSM& \textbf{18.81} & \textbf{13.02} & \textbf{10.74} & \textbf{9.63} & 100.00$^*$& 100.00$^*$& 28.94& \textbf{25.14} & 19.73& 17.26\\
& MI-FGSM & 12.48& 10.98& 9.17 & 7.18 & 100.00$^*$& 99.34$^*$& 33.73& 26.33& 22.24& \textbf{17.87} \\\midrule
\textbf{S2T-medium} & PGD \cite{olivier2022there} & 11.04& 10.23& 8.01 & 7.27 & 37.68& 29.46& 100.00$^*$& 100.00$^*$& 21.25& 21.98\\
& SAGO & \textbf{15.28} & \textbf{11.92} & \textbf{11.54} & \textbf{8.41} & \textbf{43.99} & 29.74& 78.64$^*$& 58.14$^*$& \textbf{34.87} & 23.31\\
& VMI-FGSM& 12.71& 11.28& 9.89 & 8.30 & 39.42& 31.11& 100.00$^*$& 69.07$^*$& 24.11& 24.22\\
\textbf{} & MI-FGSM & 14.10& 11.13& 10.43& 8.35 & 42.00& \textbf{31.20} & 100.00$^*$& 93.15$^*$& 26.09& \textbf{24.40} \\\midrule
\textbf{S2T-large} & PGD \cite{olivier2022there} & 15.01& 13.04& 11.10& 8.92 & 44.51& 32.75& 47.69& 35.33& 100.00$^*$& 100.00$^*$\\
& SAGO & \textbf{20.71} & \textbf{14.64} & \textbf{16.87} & \textbf{11.07} & \textbf{54.86} & 30.75& \textbf{52.38} & 33.63& 76.29$^*$& 59.31$^*$\\
& VMI-FGSM& 17.60& 13.31& 13.84& 10.66& 47.86& 33.24& 50.74& 35.36& 100.00$^*$& 95.44$^*$\\
& MI-FGSM & 19.04& 13.80& 14.44& 10.53& 49.05& \textbf{35.17} & 51.58& \textbf{37.67} & 100.00$^*$& 100.00$^*$ \\
\bottomrule \bottomrule 
\end{tabular}}
\vspace{-0.2cm}
\end{table*}

\vspace{-0.2cm}
\section{Transferable Adversarial Attacks}
In this section, we initiate by defining the research problem and then proceed to outline the proposed approaches.
\vspace{-0.3cm}
\subsection{Problem Formulation}
\vspace{-0.1cm}
Our goal is to discover the optimal perturbation $\Delta$ that not only degrades the recognition performance of a target ASR model but also remains indistinguishable from human listeners by adhering that the perturbation must be smaller than the allowed constrain. Therefore, our objective can be formulated as a constrained optimization problem:
\vspace{-0.2cm}
\begin{equation}
\vspace{-0.2cm}
 \Delta = \arg \min_{\epsilon} \mathcal{L}_{adv}(h(x + \epsilon),y) \quad \text{s.t.} \quad ||\epsilon||_{p} \leq \xi,
 \label{eq: objective}
\vspace{-0.1cm}
\end{equation}
where \( x \) is the original audio input to the ASR system \( h(.) \) and \( y \) is the ground-truth text label. The perturbation \( \epsilon \) is constrained by \( \ell_p \)-norm to ensure audio imperceptibility and \( \xi \) is the maximum adversarial perturbation allowed.
$\mathcal{L}_{\text{adv}}$ represents the negative value of the cross entropy loss between $y$ and predicted tokens from ASR.
\vspace{-0.3cm}
\subsection{Transferable Attack via Gradient Optimization}
\begin{figure}[t]
\vspace{-0.4cm}
\centering
\includegraphics[width=87mm]{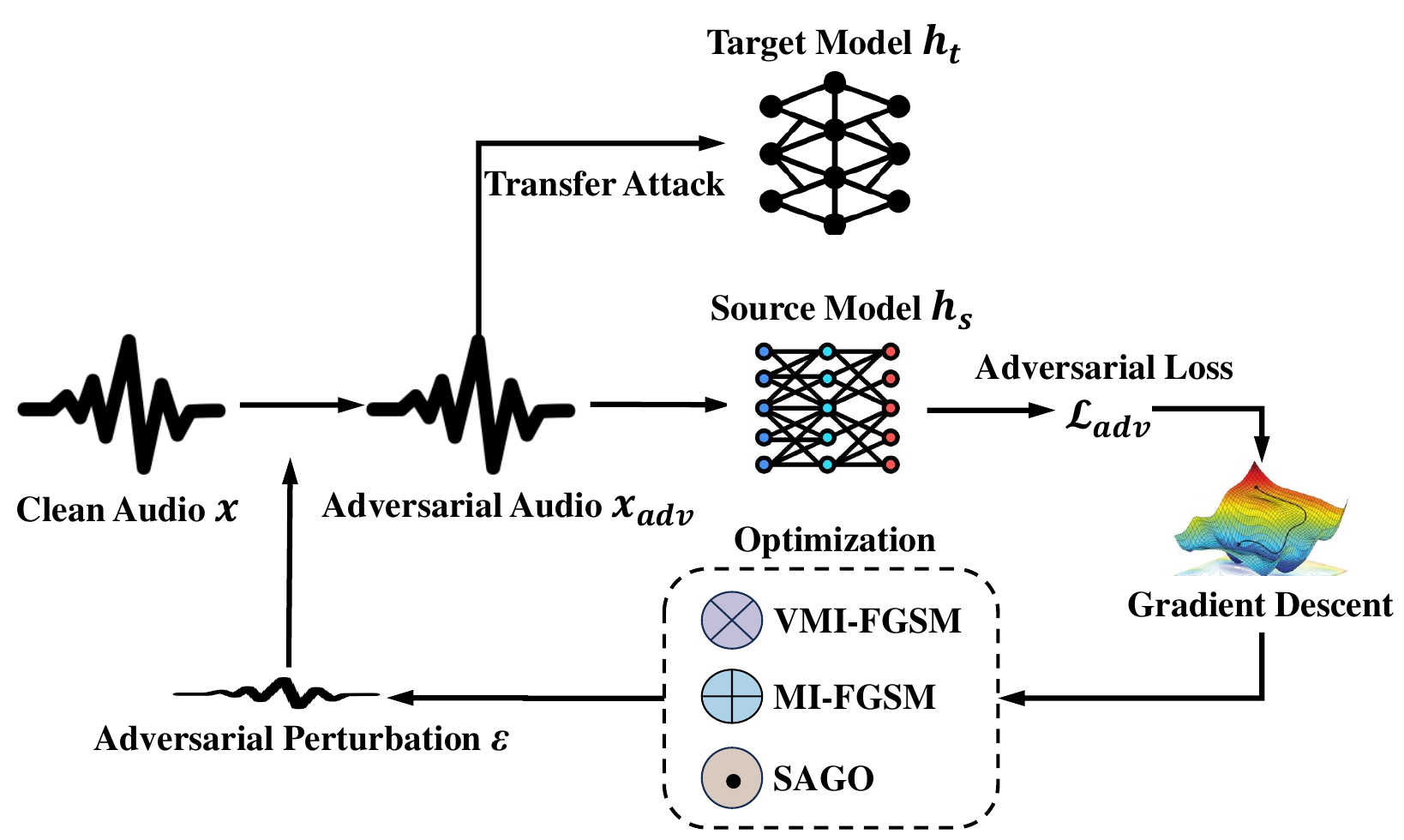}
\vspace{-0.5cm}
\caption{The overview network architecture of the proposed three transferable attacks (1) Speech-aware gradient optimization (SAGO) (2) MI-FGSM and (3) VMI-FGSM.}
\label{proposeFig}
\vspace{-0.5cm}
\end{figure}

Instead of relying on obtaining full access to the details of the target ASR model, we propose to conduct transferable adversarial attacks which only requires the output information of black-box target model.
This involves attacking source ASR model \( \textit{\textbf{h}}_\textit{{\textbf{s}}} \) and generating adversarial samples \( x_{\text{adv}} \) to serve as input for the target black-box model \( \textit{\textbf{h}}_\textit{{\textbf{t}}} \), using three novel gradient optimization approaches, as depicted in Fig~\ref{proposeFig}. We aim to create transferable attacks that serve as crucial surrogates for accessing the robustness of target ASR models (i.e., Whisper \cite{olivier2022there} and Speech2text \cite{wang2020fairseq}) before they are deployed. Each approach is elaborated below.

\subsubsection{Speech-Aware Gradient Optimization}
We propose Speech-Aware Gradient Optimization (SAGO) as a transferable attack approach for ASR systems, incorporating iterative speech-aware gradient optimization tailored to the objectives of attacking ASR models effectively. 
Specifically, SAGO is defined by iterative updating perturbations through Eq.~\ref{eq: metatrain} using speech-aware gradients in relation to the loss function $\mathcal{L}_{\text{adv}}$.
For each iteration, the updated perturbation $\epsilon$ is computed as:
\vspace{-0.2cm}
\begin{equation}
\vspace{-0.2cm}
\epsilon \leftarrow \Pi \left\{ \epsilon + \alpha \cdot sign (\textbf{M} \odot \bigtriangledown_{\epsilon} \mathcal{L}_{adv}(f(x + \epsilon), y)) \right \},
\label{eq: metatrain}
\vspace{-0.1cm}
\end{equation}
where \( \alpha \) is the learning rate, \( \bigtriangledown_{\epsilon} \mathcal{L}_{\text{adv}} \) is the gradient of the adversarial loss with respect to the perturbation \( \epsilon \).
During initialization of the attack, we compute the VAD mask $\textbf{M}$ and use it to filter out the gradient associated with non-speech components based on such mask, as shown in Fig~\ref{proposeFig}. 
$\odot$ implies element-wise product. \( \Pi \{\cdot\} \) is the projection function that enforces the \( \ell_p \) constraint on the perturbation:
\begin{equation}
\vspace{-0.2cm}
\Pi_{\ell_p}(x_{\text{adv}}) = \arg\min_{z \in S} \| x_{\text{adv}} - z \|_p,
\vspace{-0.1cm}
\end{equation}
where \( z \) is an element within the feasible set \( S \), and \( \| \cdot \|_p \) denotes the \( \ell_p \) norm. \( \Pi \{\cdot\} \) maps \( x_{\text{adv}} \) to the closest point \( z \) in \( S \) under the \( \ell_p \) norm.
SAGO integrates speech-aware knowledge to general projected gradient descent (PGD)~\cite{madry2017towards,olivier2022there} attack for gradient optimization, thereby emphasizing the significance of attacking speech segments instead of non-speech parts within an utterance for the purpose of recognition.  SAGO leverages the difference in how the human brain and ASR systems process speech by discarding non-speech components in an audio sample, which humans are primed to ignore.

Incorporating the VAD mask not only makes our attack more targeted but also potentially increases its transferability. By focusing on the speech parts crucial to ASR, the generated adversarial examples are more likely to be effective across different ASR models, which may have varying sensitivities to speech and non-speech components.

\subsubsection{Momentum Based Gradient Optimization}

Inspired by the effective transferable adversarial attack approaches on image classification with gradient optimization~\cite{dong2018boosting,wang2021enhancing}, we propose to employ the following momentum-based transferable adversarial attack methods, momentum iterative fast gradient sign method (MI-FGSM) and variance tuning momentum iterative fast gradient sign method (VMI-FGSM), for attacking black-box ASR models. 

\textbf{MI-FGSM}: By incorporating a momentum term into ASR attacks, our MI-FGSM aims to stabilize update directions, aiding in escaping from local minima during iterations, and yielding more transferable adversarial examples. For each iteration, the updated perturbation $\epsilon$ can be computed by:
\begin{equation}
\vspace{-0.1cm}
\epsilon \leftarrow \Pi \left\{ \epsilon + \alpha \cdot sign(\bigtriangledown_{\epsilon} \mathcal{L}_{adv}(f(x + \epsilon), y) + \textbf{G}) \right\},
\label{eq: mi-fgsm}
\vspace{-0.1cm}
\end{equation}
where \( \textbf{G} \) represents the momentum term, serving as a velocity vector in the gradient direction of $\mathcal{L}_{\text{adv}}$ across iterations. Momentum is expected to help remember previous gradients and navigate through narrow valleys, small humps, and unfavorable local minima \cite{duch1998optimization,dong2018boosting}.

\textbf{VMI-FGSM}: In addition to MI-FGSM, our VIM-FGSM additionally tunes the current gradient with the gradient variance in the neighborhood of the previous data point to stabilize the update direction and escape from poor local optima. For each iteration, the updated perturbation $\epsilon$ can be computed by:
\begin{equation}
\vspace{-0.1cm}
\epsilon \leftarrow \Pi \left\{ \epsilon + \alpha \cdot sign(V(\bigtriangledown_{\epsilon} \mathcal{L}_{adv}(f(x + \epsilon), y)) + \textbf{G}) \right\}
\label{eq: vmi-fgsm}
\vspace{-0.1cm}
\end{equation}
where \( V(\cdot) \) denotes a function used for additional variance tuning to reduce the variance of the gradient at each
iteration, thereby stabilizing the gradient update direction.
Our SAGO, MI-FGSM and VMI-FGSM apply fast gradient for N iterations. 

\vspace{-0.6cm}
\subsection{Time-Domain Attack}
\vspace{-0.1cm}
To craft adversarial audios flexibly and get rid of the traditional non-derivative feature processing \cite{haque2023slothspeech,povey2011kaldi,watanabe2018espnet,abdullah2021demystifying}, we propose to use a differentiable feature extractor to obtain 80-dimension log filterback features for three proposed attacks as ASR model input features. This enables gradients to back-propagate all the way to the input audio through the feature extractor, facilitating a time-domain attack. By employing these time-domain attack methods, we aim to enhance the transferability of adversarial attacks across different ASR models.

\begin{table*}[t]
\vspace{-1.2cm}
\caption{Comparison of the recognition performance (WER\%) of attack approaches under librispeech dataset. Source refers to the white-box source ASR model, while target pertains to the black-box victim ASR model. Clean and noise denote the recognition performance when decoding original clean speech and noisy speech samples, respectively. * denotes white-box attacks.}
\resizebox{\textwidth}{!}{
\begin{tabular}{ll|rrrrrrrrrr}
\toprule \midrule
\multicolumn{1}{c}{} & \textbf{Target Models} & \multicolumn{2}{c}{\textbf{Whisper-tiny}} & \multicolumn{2}{c}{\textbf{Whisper-base}} & \multicolumn{2}{c}{\textbf{S2T-small}} & \multicolumn{2}{c}{\textbf{S2T-medium}}& \multicolumn{2}{c}{\textbf{S2T-large}} \\\midrule
\multicolumn{1}{c}{\textbf{Source Models}} & \textbf{SNR}& \multicolumn{1}{c}{\textbf{30}} & \multicolumn{1}{c}{\textbf{35}} & \multicolumn{1}{c}{\textbf{30}} & \multicolumn{1}{c}{\textbf{35}} & \multicolumn{1}{c}{\textbf{30}} & \multicolumn{1}{c}{\textbf{35}} & \multicolumn{1}{c}{\textbf{30}} & \multicolumn{1}{c}{\textbf{35}} & \multicolumn{1}{c}{\textbf{30}} & \multicolumn{1}{c}{\textbf{35}} \\\midrule
 & Clean & \multicolumn{2}{c}{5.95} & \multicolumn{2}{c}{4.45}& \multicolumn{2}{c}{3.72}& \multicolumn{2}{c}{3.34}& \multicolumn{2}{c}{3.08}\\
\textbf{} & White Noise \cite{olivier2022there}& \multicolumn{1}{r}{6.14} & \multicolumn{1}{r}{6.06} & \multicolumn{1}{r}{4.49} & \multicolumn{1}{r}{4.48} & \multicolumn{1}{r}{3.72} & \multicolumn{1}{r}{3.85} & \multicolumn{1}{r}{3.31} & \multicolumn{1}{r}{3.39} & \multicolumn{1}{r}{3.07} & \multicolumn{1}{r}{3.24} \\
\textbf{} & Babble Noise \cite{shi2022robust}& \multicolumn{1}{r}{6.14} & \multicolumn{1}{r}{6.25} & \multicolumn{1}{r}{4.49} & \multicolumn{1}{r}{4.48} & \multicolumn{1}{r}{3.72} & \multicolumn{1}{r}{3.88} & \multicolumn{1}{r}{3.31} & \multicolumn{1}{r}{3.15} & \multicolumn{1}{r}{3.07} & \multicolumn{1}{r}{3.03} \\
\textbf{} & Music Noise \cite{shi2022robust}& \multicolumn{1}{r}{6.25} & \multicolumn{1}{r}{6.13} & \multicolumn{1}{r}{4.44} & \multicolumn{1}{r}{4.44} & \multicolumn{1}{r}{3.86} & \multicolumn{1}{r}{3.79} & \multicolumn{1}{r}{3.33} & \multicolumn{1}{r}{3.23} & \multicolumn{1}{r}{3.10} & \multicolumn{1}{r}{3.11} \\
\textbf{} & Natural Noise \cite{shi2022robust}& \multicolumn{1}{r}{6.15} & \multicolumn{1}{r}{6.17} & \multicolumn{1}{r}{4.46} & \multicolumn{1}{r}{4.48} & \multicolumn{1}{r}{3.83} & \multicolumn{1}{r}{4.05} & \multicolumn{1}{r}{3.24} & \multicolumn{1}{r}{3.30} & \multicolumn{1}{r}{3.11} & \multicolumn{1}{r}{3.11} \\
\midrule
\textbf{Whisper-tiny} & PGD \cite{olivier2022there}& 92.34$^*$& 95.30$^*$ & 43.77& 31.8 & 21.01& 13.68& 15.88& 11.64& 13.59& 9.75 \\
& SAGO & 100.00$^*$& 91.35$^*$& 48.14& 34.37 & 27.06& \textbf{19.94} & 20.48& \textbf{15.57} & \textbf{18.73} & 12.34\\
& VMI-FGSM& 97.83$^*$& 85.15$^*$& 47.26& 33.69& 24.89& 16.60& 19.28& 13.64& 17.23& 12.17\\
& MI-FGSM & 98.19$^*$& 78.37$^*$& \textbf{56.03} & \textbf{34.86} & \textbf{27.39} & 17.49 & \textbf{20.91} & 14.08 & 18.26 & \textbf{12.35} \\\midrule
\textbf{Whisper-base} & PGD \cite{olivier2022there} & 53.46& 37.51& 82.74$^*$& 71.86$^*$& 20.85& 14.96& 17.46& 12.71& 15.79& 11.29\\
& SAGO & 51.21& \textbf{41.73} & 83.89$^*$& 70.98$^*$& \textbf{28.30} & \textbf{20.06} & \textbf{21.83} & \textbf{14.98} & \textbf{19.49} & \textbf{13.70} \\
& VMI-FGSM& 51.14& 39.63 & 81.57$^*$& 69.60$^*$& 23.36& 16.33& 20.33& 14.46& 18.87 & 13.19 \\
& MI-FGSM & \textbf{53.82} & 38.10& 83.89$^*$& 70.98$^*$& 24.36 & 16.78 & 21.12 & 14.90 & 18.96& 13.34\\\midrule
\textbf{S2T-small} & PGD \cite{olivier2022there} & 12.56& 10.56& 8.82 & 8.03 & 100.00$^*$ & 100.00$^*$& 26.69& 22.43& 17.14& 15.23\\
& SAGO & \textbf{24.91} & \textbf{18.30} & \textbf{19.52} & \textbf{13.80} & 100.00$^*$& 76.79$^*$& \textbf{49.30} & \textbf{37.57} & \textbf{37.49} & \textbf{24.96} \\
& VMI-FGSM& 12.18 &9.79 & 8.68 & 7.05 & 100.00$^*$& 86.77$^*$& 31.11& 24.63 & 20.94& 16.74\\
& MI-FGSM & 15.92& 15.04& 12.12& 9.75 & 100.00$^*$& 86.77$^*$& 32.39& 26.84& 22.57& 19.20 \\\midrule
\textbf{S2T-medium} & PGD \cite{olivier2022there} & 13.09& 11.80& 9.68 & 8.44 & 39.52& 30.93& 100.00$^*$& 100.00$^*$& 23.98& 21.25\\
& SAGO & \textbf{28.41} & \textbf{20.34} & \textbf{24.35} & \textbf{16.03} & \textbf{65.93} & \textbf{45.31} & 88.90$^*$& 70.35$^*$& \textbf{50.98} & \textbf{34.87} \\
& VMI-FGSM& 15.97& 13.89& 12.02& 10.76& 40.92& 33.89& 100.00$^*$& 81.04$^*$& 27.95& 24.11\\
\textbf{} & MI-FGSM & 17.54& 14.74& 13.72& 11.15& 41.99& 34.13 & 100.00$^*$& 79.22$^*$& 30.35& 26.09 \\\midrule
\textbf{S2T-large} & PGD \cite{olivier2022there} & 18.11& 14.32& 14.21& 11.09& 49.65& 35.59& 46.65& 35.59& 100.00$^*$& 79.63$^*$\\
& SAGO & \textbf{36.05} & \textbf{22.20} & \textbf{29.73} & \textbf{19.29} & \textbf{59.32} & \textbf{42.67} & \textbf{62.37} & \textbf{44.30} & 84.22$^*$& 63.44$^*$\\
& VMI-FGSM& 22.66& 21.07& 17.50& 13.94& 49.65& 37.63& 50.21& 39.50& 92.23$^*$& 68.65$^*$\\ 
& MI-FGSM & 27.42& 17.45& 19.54& 14.74& 49.42& 37.53 & 50.74& 38.67 & 86.99$^*$& 67.50$^*$ \\
\bottomrule \bottomrule
\end{tabular}}
\label{table2}
\vspace{-0.4cm}
\end{table*}

\vspace{-0.5cm}
\section{Experiments}
\vspace{-0.2cm}
\subsection{Datasets}
\vspace{-0.1cm}
We employ two widely recognized English datasets from huggingface: the LibriSpeech dataset \cite{panayotov2015librispeech} and the LJ-Speech dataset \cite{ljspeech17}. We assess the first 500 utterances from the LJ-Speech and the first 500 utterances of the clean test subset from the LibriSpeech for evaluation. All audios are re-sampled to 16,000 Hz.
 \vspace{-0.4cm}
\subsection{Experimental Setups}
 \vspace{-0.1cm}
\textbf{Models}: we utilize SOTA ASR frameworks: Whisper \cite{radford2023robust} and Speech2text (S2T) \cite{wang2020fairseq} as both source and target ASR models. Whisper employs weakly supervised speech recognition using transformer with 680,000 hours of data~\cite{radford2023robust}, while S2T is a transformer-based seq2seq (encoder-decoder) model designed for E2E ASR~\cite{wang2020fairseq}. Our experimentation involves five ASR models: Whisper-tiny, Whisper-base, S2T-small, S2T-medium, and S2T-large, for transfer attacks between them.

\textbf{Baselines}: we employ PGD \cite{olivier2022there} as a strong baseline (re-implement \cite{olivier2022there} to allow more ASR models). White noise is considered as a baseline \cite{olivier2022there} for comparison. We also propose to investigate three more widely used noise categories as in \cite{shi2022robust} that include babble, music and natural noises from MUSAN dataset \cite{snyder2015musan}. All noises are augmented to the input clean audios for ASR model decoding.

\textbf{Attack Setups}: for all attacks, p is set to $\infty$, and attack iteration steps N is set to 50. Perturbation size $\xi$ is set to 0.002 and 0.0035, representing average signal-to-noise ratios (SNRs) of 35dB and 30dB, respectively, indicative of exceptionally low levels of noise. VAD is implemented by using a cepstral power measurement to detect speech and trim non-speech parts from the front and end of the audio. 

\textbf{Evaluation Method}: to assess the degradation of ASR performance, we measure it in terms of Word Error Rate (WER), which calculates the ratio of total insertions, substitutions, and deletions to the total number of words. We set a maximum WER value of 100.00 \% for values exceeding 100.00 $\%$. A higher WER signifies poorer ASR performance and a more successful attack.
\vspace{-0.2cm}
\section{Results and Discussion}
\vspace{-0.1cm}
We study the performances of white-box and transferable attacks, the effect of noise sensitivity and human imperceptibility. Table~\ref{table1} and \ref{table2} summarize the recognition performance of both baseline and our proposed methods at differnt SNR values for the LJ-speech and Librispeech datasets, respectively. 

\vspace{-0.4cm}
\subsection{White-box Attacks}

We build white-box ASR attackers to serve as the source models for conducting black-box attacks. We initially observe from Table~\ref{table1} and \ref{table2} (*) that our proposed attacks surpass noise baselines and attain good performance (WER$>$50 \%) together with strong baseline PGD. Note that PGD is a strong method for white-box attacks for general scenarios as indicated in~\cite{madry2017towards}.

\vspace{-0.4cm}
\subsection{Noise Sensitivity}
\vspace{-0.2cm}

We observe that the inclusion of noises like babble, music and natural sounds and white noises does not degrade ASR model performances and perform worse than our proposed attacks. 
This underscores the varying vulnerability of ASR models' to additive noises and transferable attacks, emphasizing the superiority of adversarial perturbations over added noises. These findings emphasize the importance of assessing ASR model robustness against transferable attacks.

\vspace{-0.4cm}
\subsection{Transferable Attacks}
\vspace{-0.2cm}
Our proposed attacks generally induce significant performance degradation, yielding higher WER compared to baselines across five models and two datasets. This highlights the effectiveness of our approaches' transferability, potentially rendering ASR models unsuitable for various practical applications. For instance, a 65\% WER performance can severely impact the quality of video captioning and car navigation systems when a source S2T-medium model attacks a target S2T-small model. For Whisper-based models, FGSM approaches tend to be more effective, indicating their robustness in scenarios with complex models that may cause gradient-based methods to struggle. For S2T models, SAGO shows better performance, likely due to its ability to target speech-specific characteristics through VAD masking. This suggests that the choice of method should be context-dependent.

\vspace{-0.3cm}
\subsection{Human Imperceptibility}
\vspace{-0.1cm}
Adversarial samples and codes can be accessed through the link~\footnote{\url{https://xiaoxue1117.github.io/TransferableASR/}}. The adversarial samples from our proposed attacks sounds similar to the original clean audio, achieving higher WER (better transferable attack performance) compared with baselines. This indicates that the adversarial samples are undetectable to human and the proposed attacks captures aspects of speech that are generally not perceived by human subjects, but that are crucial for ASR accuracy.
\vspace{-0.3cm}
\section{Conclusion}
\vspace{-0.1cm}
The proposed SAGO successfully empowers attackers to harness speech-aware insights from VAD for transferable ASR attacks.
Our proposed MI-FGSM and VMI-FGSM also serves as an important step into the comprehensive exploration of transferable attack against ASR, where we propose to leverage momentum-oriented gradient optimization and demonstrate their effectiveness as time-domain attacks with a differentiable feature extractor design.
The proposed attacks not only demonstrate their cross-model transferable effectiveness but also affirm that the generated audio does not impact human understandability. The combination of the proposed methods will be studied further in future work.

\footnotesize
\bibliographystyle{IEEEtran}
\bibliography{mybib}

\end{document}